\documentclass[english]{emulateapj}

\slugcomment{Astrophysical Journal 699 (2009) 1252-1260.}

\shorttitle{Cosmological Voids in the SDSS}

\shortauthors{Foster \& Nelson}

\makeatletter

\usepackage{babel}

\begin{document}

\title{THE SIZE, SHAPE AND ORIENTATION OF COSMOLOGICAL VOIDS IN THE SLOAN DIGITAL
SKY SURVEY}

\author{Caroline Foster$^{1,2}$ and Lorne A. Nelson$^1$}

\affil{$^1$ Physics Department, Bishop's University, 2600 College Street, Sherbrooke, Quebec Canada J1M 1Z7\\
$^2$Centre for Astrophysics \& Supercomputing, Swinburne University, Hawthorn, VIC 3122, Australia}

\email{cfoster@astro.swin.edu.au}

\begin{abstract}
We present a detailed description of our void finding algorithm which is an extension of the prescription by Hoyle and Vogeley (2002). We include a discussion of the reproducibility and robustness of the algorithm as well as the statistical significance of the detected voids. We apply our void finder to the Data Release 5 (DR5) of the Sloan Digital Sky Survey (SDSS) and identify 232 cosmological voids. A void catalog which contains the most salient properties of the detected voids is created. We present a statistical analysis of the distribution of the size, shape and orientation of our identified cosmological voids. We also investigate possible trends with redshift for $0.04\le z \le0.16$. We compare our results to those from an identical analysis of a mock catalog based on the $\Lambda$CDM model and find reasonable agreement. However, some statistically significant differences in the overall orientation of cosmological voids are present and will have to be reconciled by further refinement of the simulations.
\end{abstract}


\section{INTRODUCTION}\label{sec:INTRODUCTION}

In the 1970s and the 1980s, large-scale redshift surveys have revealed the bubbly structures in our universe on a megaparsec (Mpc) distance scale. Although some degree of clustering was expected, the discovery of large-scale structures was a surprise (Fairall 1998). Cosmological voids are large regions that are nearly devoid of luminous matter and are surrounded by walls, filaments and clusters of galaxies. As was first proposed by Zel'dovich (1970), they are now best explained as the result of gravitational instabilities in the early universe. Because of their enormous size, cosmological voids must have formed early in the history of the universe and evolved to their present state as expanding low-density regions as predicted by the cold dark matter models with cosmological constant ($\Lambda$CDM).

Significant improvements in computational power have made it possible to model the formation and evolution of large-scale structures through N-body simulations (an example being the Millennium Simulation, see Springel et al. 2005). Studies of the distribution of sizes of voids in the observed galaxy distribution have been performed (see for example, Plionis \& Basilakos 2002; Muller et al. 2000; Muller \& Maulbetsch 2004). An analysis of void properties with respect to CDM models and the effects of galaxy/dark matter biasing has been carried out by several authors (e.g., Mathis \& White 2002; Benson et al. 2003; Tikhonov \& Klypin 2009; Tinker \& Conroy 2009). Indirect evidence of the size evolution of cosmological voids has been provided by Conroy et al. (2005) who find that the observed void probability function\footnote{The void probability function (VPF) is a measure of the probability that a sphere of a given size is devoid of galaxies in a given survey.} at low redshift ($z\approx0$) as measured in the Sloan Digital Sky Survey (SDSS) and high redshift ($z\approx1$) as measured in the DEEP2 Redshift Survey is consistent with that expected within the $\Lambda$CDM framework.

With the increasing amount of observational data now becoming available, it is important to investigate the distribution of the properties of cosmological voids. Indeed, the fifth data release of SDSS (Adelman-McCarthy et al. 2007) contains sufficient data to be able to select statistically homogeneous galaxy samples out to a redshift of about 0.16. It is thus possible to look at the overall distribution of sizes, shapes and orientations of voids.

Moreover, Ryden \& Mellot (1996) proposed a method of verifying the evolution of voids by studying redshift-space distortions via the shape and orientation of cosmological voids in redshift space. The redshift is a measure of the speed of recession of an object with respect to the line-of-sight. For this reason, peculiar velocities (e.g., random velocity dispersions in virialized structures and coherent infall), large-scale cosmological distortion as well as the expansion of the universe all have an impact on the measured redshift. Unfortunately, these factors are hard to deconvolve.

The basic idea is the following: If a hypothetical spherical void with negligible galaxy velocity dispersion around its boundary is expanding in real space axisymmetrically, it will appear as an elongated ellipse with its semi-major axis aligned with the line-of-sight in redshift space. On the other hand, if such an idealized void is collapsing in real space, its semi-major axis will intersect the line-of-sight at $90^{\circ}$ in redshift space. Thus, under the assumptions that 1) the intrinsic shape of the majority of voids is on average spherical in real space, and 2) the galaxies on the boundary of voids have negligible velocity dispersion, then if most voids in redshift space are found to intersect the line-of-sight at angles smaller than the average expected from a random distribution in three dimensions (i.e. $\left\langle \phi\right\rangle < 57.3^{\circ}$), it would indicate that voids are expanding in real space on average. Otherwise, if cosmological voids in astronomical data are found to intersect the line-of-sight at angles $\left\langle\phi\right\rangle > 57.3^{\circ}$, it would suggest that most voids are collapsing in real space. Of course, the effects of other redshift-space distortions on the distribution of cosmological voids such as 'fingers of God' and the more subtle Kaiser effect (Kaiser 1987) caused by the coherent infall towards/away from over-densities/under-densities can play an important role. For example, fingers of God stretch inside the voids thereby reducing their size, and changing their shape as well as their orientation. As shown by Ryden \& Mellot (1996), the net redshift-space distortions of cosmological voids (i.e. elongation or compression along the line-of-sight) in cosmological simulations depends on the initial conditions and cosmological parameters (e.g. the slope of the power-spectrum and $\Omega_{tot}$). This suggests that a comparison of the shape and orientation of cosmological voids in observational data with the predictions of semi-analytic models could be a useful test of the cosmological and semi-analytic models' input parameters.

In order to do this, one must find an objective and accurate way of identifying and quantifying the properties of voids. Several void finding algorithms have been created in order to fulfill these objectives. There are a wide variety of definitions and techniques that have been employed by various groups. Only recently has a serious attempt been made to compare and contrast these different algorithms (see Colberg et al. 2008, and references therein for an overview of the different void finding algorithms in use, including the present). We examined several algorithms and selected the one that we felt was best suited to a robust analysis of the distribution of sizes, shapes and orientation of cosmological voids. We adopted an algorithm that objectively identifies voids in the observed galaxy distribution in a way that best reproduces the results of a visual inspection. Our algorithm is an adaptation and extension of the algorithm by Hoyle \& Vogeley (2002, hereafter H\&V) which has been extensively employed to identify voids and study their properties as well as those of their constituent galaxies in several redshift surveys and structure formation simulations (Hoyle \& Vogeley 2004; Hoyle et al. 2005).

This paper is divided as follows: \S II describes our void finding algorithm, while \S III, \S IV and \S V treat the reproducibility of our results, robustness of the algorithm and statistical significance of voids, respectively. \S VI gives a brief description our Sloan Digital Sky Survey sample and \S VII presents the properties of cosmological voids in our sample and a comparison to a mock catalog. Our conclusions are given in \S VIII.

\section{VOID FINDING ALGORITHM}

The void finding algorithm is based on the prescription of Hoyle \& Vogeley (see H\&V, 2004; Hoyle et al. 2005). It has been thoroughly tested on the Updated Zwicky Catalog (UZC) and the Point Source Catalog Redshift survey (PSCz). The algorithm is subdivided into seven distinct steps: (i) data input; (ii) classification of galaxies as field or wall galaxies; (iii) detection of the empty cells in the distribution of wall galaxies; (iv) growth of the maximal sphere; (v) classification of the unique voids; (vi) enhancement of the void volume; and (vii) calculation of the void properties. Below is a detailed description of each step.

\subsection{Data Input}

Once the dataset has been selected, equatorial coordinates are converted to comoving Cartesian coordinates $\left(X_{c},\, Y_{c},\, Z_{c}\right)$. This is carried out using the following formulae:

\begin{equation}
X_{c}=D_{c}\sin\left(\frac{\pi}{180^{\circ}}(90^{\circ}-\delta)\right)\cos\left(\frac{\pi\alpha}{180^{\circ}}\right)\end{equation}

\begin{equation}
Y_{c}=D_{c}\sin\left(\frac{\pi}{180^{\circ}}(90^{\circ}-\delta)\right)\sin\left(\frac{\pi\alpha}{180^{\circ}}\right)\end{equation}

\begin{equation}
Z_{c}=D_{c}\cos\left(\frac{\pi}{180^{\circ}}(90^{\circ}-\delta)\right)\end{equation}

\noindent where $\alpha$ is the right ascension and $\delta$ is the declination (both in degrees). Also note that $D_{c}$ is the comoving distance in megaparsecs (Mpc) given by the following formula (Hogg 2000):

\begin{equation}
D_{c}=\frac{c}{H_{0}}\int_{0}^{z}\frac{dz'}{\sqrt{\Omega_{m}(1+z')^{3}+\Omega_{k}(1+z'){}^{2}+\Omega_{\Lambda}}},\label{eq:comovingdist}\end{equation}

\noindent  where $z$ is the redshift. We have adopted the following values for each of the parameters: $H_{0}=100h$, $\Omega_{m}=0.28$, $\Omega_{k}=0$, and $\Omega_{\Lambda}=1-\Omega_{m}-\Omega_{k}=0.72$. These values were chosen based on the results of WMAP (Spergel et al. 2007).

\subsection{Classification of galaxies as field or wall galaxies}

From the data, the average distance to the third nearest galaxy ($D_{3}$) as well as the standard deviation ($\sigma_{3}$) of that value are computed. The parameter $R_{3}$ is defined such that \begin{equation} R_{3}\equiv D_{3}+\lambda\sigma_{3},\end{equation} where $\lambda$ is an adjustable dimensionless parameter that together with the threshold parameter ($\xi$) most strongly affects the inferred number and sizes of voids. The threshold parameter is simply defined as the minimum allowed radius of a void. After a thorough analysis of the robustness of the algorithm using the Updated Zwicky Catalog, we chose values of $\lambda=2.0$, and $\xi=12h^{-1}$Mpc. These values of $\lambda$ and $\xi$ yield the smallest variation in the number of voids found with respect to the other parameters (see Section 4). Moreover, relatively large values of $\xi$ guarantee that all voids that are identified by the algorithm are legitimate cosmological voids as opposed to spurious voids.

Wall galaxies are then defined as those whose third nearest neighbor is closer than $R_{3}$. All other galaxies are field galaxies, and are the only ones allowed to exist inside a void.

\subsection{Detection of the empty cells in the distribution of wall galaxies}

The wall galaxies are placed in a grid whose basic cell is cubical and has a side of length $R_{3}/2$. Dividing the cell size by two yields $2^{3}$ times more holes than those used by Hoyle and Vogeley, and thus it allows us to sample the volume more precisely. The entire list of wall galaxies is sorted and the cells that contain no wall galaxies are identified. The center of each empty cell $C_{1}\equiv\left(X_{hole1},\, Y_{hole1},\, Z_{hole1}\right)$ is recorded for later use.

\subsection{Growth of the maximal sphere}

The position $G_{1}\equiv\left(X_{gal1},Y_{gal1},Z_{gal1}\right)$ of the nearest galaxy to the center of every empty cell is recorded. An empty sphere is defined as having the center of the empty cell as its center and a radius constrained by the nearest galaxy located on its surface.

A first growth vector, $\overrightarrow{v_{1}}$, pointing from the nearest galaxy to the center of the empty sphere is then computed and the radius is effectively increased in the direction of $\overrightarrow{v_{1}}$ until another galaxy is found to intersect the surface. The algorithm goes through the entire list of galaxies and finds the galaxy $G_{2}\equiv\left(X_{gal2},Y_{gal2},Z_{gal2}\right)$ which yields the smallest sphere whose center $C_{2}\equiv\left(X_{hole2},Y_{hole2},Z_{hole2}\right)$ has moved along $\overrightarrow{v_{1}}$ such that $G_{1}$ and $G_{2}$ both lie on the surface of the sphere.

A second growth vector, $\overrightarrow{v_{2}}$, is defined from the midpoint between $G_{1}$ and $G_{2}$ to the second center $C_{2}$. In a similar fashion to the first growth, the radius is effectively increased by moving the center in the direction of $\overrightarrow{v_{2}}$ until a third galaxy intersects the surface. The center $C_{3}\equiv\left(X_{hole3},Y_{hole3},Z_{hole3}\right)$ of the fully grown sphere together with its final radius constitute a hole.

The algorithm then checks that each hole lies within the boundaries of the survey. Because legitimate holes (and voids) may exist near the edge and reach beyond the survey boundaries, we allow for part of the hole to exist outside the boundary by extending it by 5-10 Mpc. Should the boundaries of the survey extend into regions where extinction due to the disk of the Milky Way is significant, then the holes along that line-of-sight are removed. The effect of extending the survey boundary (Extension) and of varying the range of galactic latitudes extinct by the disk of the Milky Way (MW width) are discussed in Section \ref{sec:ROBUSTNESS}. 

\subsection{Classification of the unique voids}

The holes are ordered from the largest to the smallest. If the radius of the first hole is larger than $\xi$, it is identified as void number one. The program then goes through the entire list of holes whose radii exceed $\xi$. If a hole overlaps any one of the constituent holes of at least two voids by more than $\beta_{3}=2\%$ of its volume, it is rejected; if it overlaps a previously defined void by at least $\beta_{1}=10\%$, it is merged with that void. Otherwise, if it does not overlap any previously defined void, it is a new void.

\subsection{Enhancement of the void volume}

The holes that have radii smaller than $\xi$ are considered next. In a similar manner as for the large holes, if a small hole overlaps constituent large holes of at least two voids by more than $\beta_{3}$, it is rejected. Otherwise, if it overlaps one and only one large hole by at least $\beta_{2}=50\%$, it is merged with the corresponding void. Expansion of the volumes of the voids by this method allows for the growth of aspherical voids.

\subsection{Calculation of the void properties}

Once the constituent holes of each void have been identified, we compute the final position, volume and equivalent spherical radius of each void using Monte Carlo methods.

\subsubsection{Determination of Shapes}

Quantifying the shape of voids is non-trivial because the agglomeration of empty spheres yields voids of complex shapes. This problem is much like the one of quantifying the shape of clusters of astronomical objects.

We have used the method of the best fit ellipsoid similar to that used by Jang-Condell \& Hernquist (2001) and Platen et al. (2007). We use Monte Carlo methods to compute the shape tensor $S$:
\begin{equation}
S_{ij}\equiv \sum_{k=1}^{N}{m_{k}r_{ki}r_{kj}} 
\label{eq1}
\end{equation}
\noindent where $r_{k}$ is the distance of the $k^{th}$ particle to the center of the void, $i$ and $j$ represent the spatial components and we assume that $m_{k}=1$ for all $k$ randomly placed particles inside the void. We then find the eigenvalues of $5S/M$ (where  $M=\sum_{k=1}^{N}{m_k}$), which are the square of the sizes of the principal axes of the best fit ellipsoid ($a^{2}$, $b^{2}$ and $c^{2}$).

We also compute $\epsilon_{1}$, $\epsilon_{2}$ and the triaxiality $T$ given by the following equations:
\begin{equation}
\epsilon_{1} = 1-\frac{b}{a},
\end{equation}
\begin{equation}
\epsilon_{2} = 1-\frac{c}{b},
\end{equation}
\begin{equation}
T = \frac{a^{2}-b^{2}}{a^{2}-c^{2}}.
\end{equation}
\noindent Thus, a prolate void in redshift space has $\epsilon_{1}>0$, $\epsilon_{2}=0$ and $T=1$. An oblate one yields $\epsilon_{1}=0$, $\epsilon_{2}>0$ and $T=0$. Finally, $\epsilon_{1}=\epsilon_{2}=0$ corresponds to a perfectly spherical void.

\subsubsection{Orientation of Voids in Redshift Space}

The eigenvectors of $5S/M$ are along the principal axis of the best fit ellipsoid corresponding to the respective eigenvalues found above. Having the vector along the semi-major axis enables us to compute the orientation (i.e., $\phi$) of the void. The orientation of a cosmological void is defined in such a way that $\phi$ is $0^{\circ}$ when the semi-major axis is aligned with the line-of-sight and $\phi=90^{\circ}$ when the semi-major axis is perpendicular to the line-of-sight.

\section{REPRODUCIBILITY: THE UPDATED ZWICKY CATALOG (UZC)}

The UZC is made up of data from both the original Zwicky Catalog (ZC) and the CfA redshift survey to $m_{Zwicky}\lesssim15.5$. At the time that the CfA survey was carried out, measuring a redshift was a tedious process. Therefore, redshifts from the previous literature were used and this led to a very inhomogeneous database. The UZC was an attempt at improving the consistency of this database and provides a revised $~2$'' accuracy set of coordinates for the objects in the ZC. The catalog contains 19,369 objects, 18,633 of which have measured redshifts for the main survey regions of $20^{\textnormal{h}}\leqslant\alpha_{1950}\leqslant4^{\textnormal{h}}$ and
$8^{\textnormal{h}}\leqslant\alpha_{1950}\leqslant17^{\textnormal{h}}$, and both with $-2.5^{\circ}\leqslant\delta_{1950}\leqslant50^{\circ}$. For more details concerning the UZC, refer to Falco et al. (1999).

For this analysis\footnote{We also carried the analysis on the Point Source Catalog redshift survey (PSCz). However, we chose not to include it in this paper as it does not contain any significantly new information.}, we used the same sample as H\&V; namely, a volume and absolute magnitude limited sample with $z_{max}=0.025$ and $M_{lim}=-18.96$. This sample contained 3500 galaxies of which 283 were field galaxies. The absolute magnitude $M$ was computed according to the formula provided in H\&V:

\begin{equation}
M=m_{Zwicky}-25-5\log\left[D_{c}\left(1+z\right)\right]-3z.
\end{equation}

\begin{figure*}
\plotone{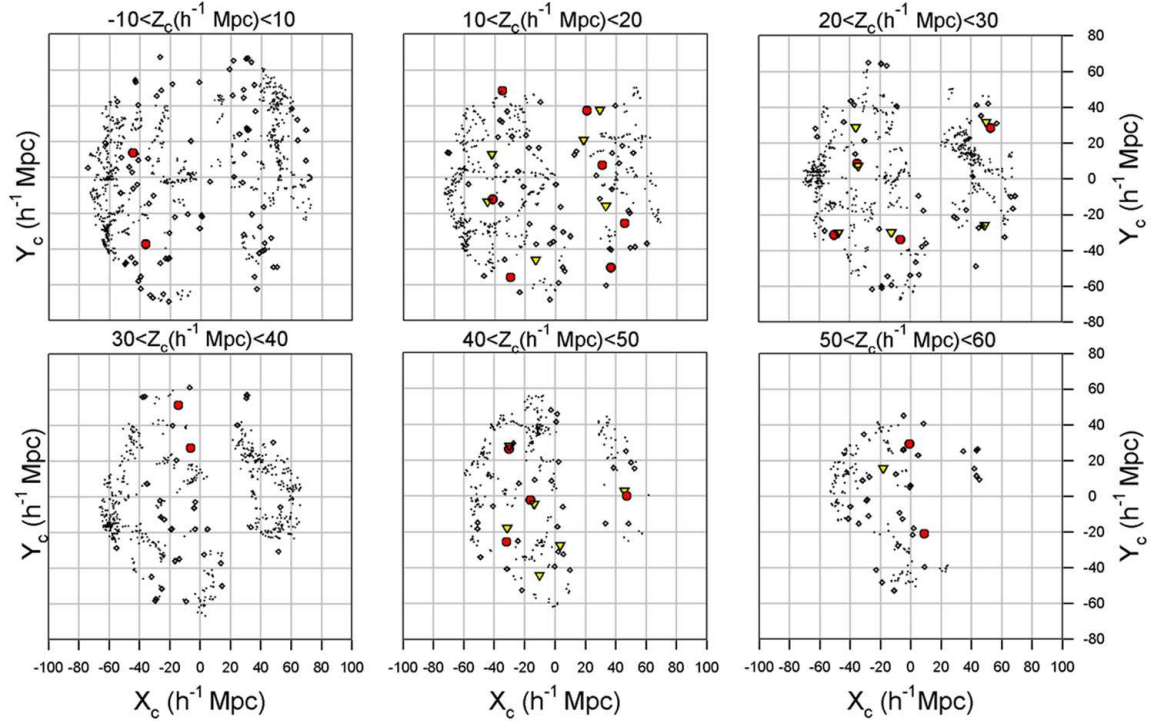} \caption{Voids in the UZC as found in our analysis and that of H\&V. The red circles and yellow triangles represent the center of the voids in the former and in the latter analyses, respectively. The black dots represent wall galaxies and the hollow diamonds represent field galaxies. The volume of the survey has been divided into six slices. The comoving Cartesian coordinates are shown and the limits on $Z_{c}$ for each slice are also noted.
\label{fig1}}
\end{figure*}

H\&V's analysis of the distribution of galaxies in the UZC revealed 19 voids. Our algorithm found 21 voids. Figure \ref{fig1} shows the position of the centers of our voids (red circles) and those of H\&V (yellow triangles). Also, in Tables \ref{tab:UZCH&V} and \ref{tab:UZC}, respectively, the results from H\&V's analysis and the present one are tabulated for comparison purposes.

The position of the center of the voids coincides at least 50\% of the time. Indeed, the two analyses are not exhaustive and should be viewed as complementary rather than contradictory. However, the sizes of voids tend to be different. Consider for example void FN1 which can be positively identified with H\&V's second void (HV2). In both cases, it is amongst the largest voids; however, HV2 has a radius 64.8\% that of FN1. We conclude that FN1 is in fact an amalgamation of HV2, HV5 and the region around $X_{c}=-10$ $h^{-1}$Mpc, $Y_{c}=-25$ $h^{-1}$Mpc, and $Z_{c}=15$ $h^{-1}$Mpc\footnote{($X_c$,$Y_c$,$Z_c$) are the cartesian comoving coordinates with origin coinciding with that of equatorial coordinates.} (see Figure \ref{fig1}). This discrepancy in the volume is probably due to the intrinsically different definitions as to what constitutes a void.

Another interesting void to consider is HV17, which has a radius that is 86.5\% that of FN20 (it contains 2 field galaxies only). In this case, the two results are very similar. Once more, our analysis tends to produce larger voids because, in general, we add more holes together. By visual inspection, one can see that both void sizes are possible. It should be noted that our voids are not always consistently larger than those of H\&V.

\begin{deluxetable}{cccccc}

\tablecaption{\label{tab:UZCH&V}Voids in the UZC according to Hoyle
\& Vogeley (2002).}

\tabletypesize{\scriptsize} \tablewidth{0pt}

\tablehead{ \colhead{HV}& \colhead{Diameter\tablenotemark{a}}&
\colhead{Volume}& \colhead{Distance}& \colhead{$\alpha$}&
\colhead{$\delta$}\\
\colhead{\#}& \colhead{($h^{-1}$Mpc)}&
\colhead{($h^{-3}$Mpc$^{3}$)}& \colhead{($h^{-1}$Mpc)}&
\colhead{(deg)}& \colhead{(deg)} }

\startdata

1& 39.73& 32828& 51.22& 197.8& 73.9\\ 2& 36.12& 24667& 42.09& 247.2&
39.8\\ 3& 33.79& 20204& 49.81& 52.1& 15.8\\ 4& 32.22& 17506& 59.56&
209.7& 52.6\\ 5& 33.92& 20443& 44.54& 168.0& 38.3\\ 6& 29.25& 13104&
49.15& 254.4& 13.7\\ 7& 28.20& 11747& 59.68& 332.5& 21.8\\ 8& 27.63&
11045& 47.74& 196.6& 12.2\\ 9& 26.75& 10025& 38.91& 334.5& 18.6\\
10& 29.63& 13625& 61.75& 136.9& 48.4\\ 11& 26.29& 9509& 45.36&
162.6& 14.7\\ 12& 23.89& 7142& 61.65& 3.7& 42.0\\ 13& 24.15& 7378&
63.02& 32.5& 19.7\\ 14& 25.54& 8723& 60.93& 256.2& 41.7\\ 15& 26.23&
9454& 51.65& 277.2& 57.5\\ 16& 26.78& 10060& 56.82& 139.4& 65.1\\
17& 23.72& 6990& 62.58& 212.2& 26.4\\ 18& 21.07& 4901& 30.35& 48.6&
21.9\\ 19& 26.91& 10203& 51.07& 141.2& 25.4

\enddata

\tablenotetext{a}{Equivalent spherical diameter.}

\end{deluxetable}

\begin{deluxetable}{cccccc}

\tablecaption{\label{tab:UZC}Voids in the UZC found using the same
selection criteria as H\&V. }

\tabletypesize{\scriptsize}

\tablewidth{0pt}

\tablehead{ \colhead{FN}& \colhead{Diameter\tablenotemark{a}}&
\colhead{Volume}& \colhead{Distance}& \colhead{$\alpha$}&
\colhead{$\delta$}\\
\colhead{\#}& \colhead{($h^{-1}$Mpc)}&
\colhead{($h^{-3}$Mpc$^{3}$)}& \colhead{($h^{-1}$Mpc)}&
\colhead{(deg)}& \colhead{(deg)} }

\startdata

1& 55.70& 90482& 40.1& 259.15& 31.61\\ 2& 50.74& 68399& 45.5& 60.61&
18.89\\ 3& 46.84& 53808& 56.1& 293.07& 66.03\\ 4& 46.24& 51767&
55.3& 331.00& 18.71\\ 5& 41.08& 36299& 63.1& 218.93& 49.87\\ 6&
37.78& 28235& 64.7& 0.09& 42.57\\ 7& 37.42& 27435& 44.3& 196.21&
14.75\\ 8& 36.66& 25797& 49.8& 187.95& 71.34\\ 9& 36.28&
25003& 60.9& 125.50& 9.85\\ 10& 35.98& 24388& 45.4& 165.81& 37.24\\
11& 35.78& 23984& 61.6& 138.80& 49.78\\ 12& 34.54& 21576& 35.5&
12.77& 26.27\\ 13& 33.46& 19614& 64.6& 28.52& 21.18\\ 14& 31.08&
15720& 63.8& 91.29& 62.71\\ 15& 29.90& 13996& 63.3& 306.32& 9.93\\
16& 28.52& 12146& 64.3& 242.21& 11.97\\ 17& 28.08& 11593& 47.0&
162.45& 9.09\\ 18& 27.42& 10794& 65.5& 211.92& 25.96\\ 19& 25.86&
9055& 51.9& 225.87& 7.36\\ 20& 25.68& 8867& 44.3& 102.61& 50.52\\
21& 25.56& 8743& 62.4& 105.34& 31.57

\enddata

\tablenotetext{a}{Equivalent spherical diameter.}

\end{deluxetable}

\section{ROBUSTNESS OF THE ALGORITHM}\label{sec:ROBUSTNESS}

There are several parameters used in the void finder algorithm that influence the number of voids found. Ideally, a void finding algorithm should find a consistent number of voids regardless of the values of the input parameters. To determine which values of the primary parameters ($\lambda$, $\xi$) should be used as the default values for subsequent void identifications, several runs were made using the UZC for which the {}``secondary parameters'' were varied. The reason these two parameters are grouped together for this analysis is because, as opposed to any of the other parameters, their influence on the number of voids identified is
closely interrelated. This was only found after considerable numerical experimentation.

\begin{figure}
\plotone{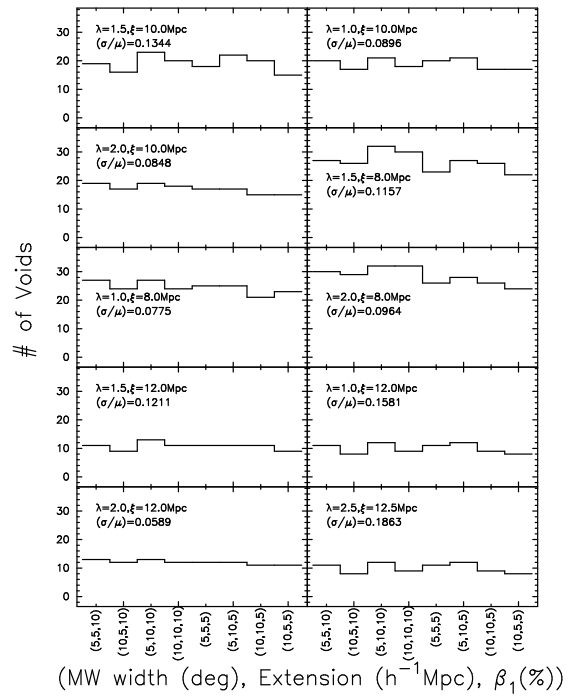} \caption{Summary of the variability analysis. The set of primary parameters which yields the smallest variability is $\lambda=2.0,\,\xi=12h^{-1}$Mpc. \label{fig2}}
\end{figure}

The {}``effective'' secondary parameters, referred to above, are: (1) the width of the disk of the Milky Way (MW width) given in degrees of galactic latitude; (2) the extension to the maximum radial boundary of the survey (Extension) given in $h^{-1}$Mpc; and, (3) $\beta_{1}$, which determines how holes are merged to form voids. The parameter $\beta_{2}$ and $\beta_{3}$ are not included in this analysis as their value does not influence the number of voids found. To compute the robustness of a particular combination ($\lambda$, $\xi$), the standard deviation for each permutation of the other parameters was calculated and it was normalized to the average number of voids ($\sigma/\mu$). This parameter is referred to as the normalized variability. As can be seen in Figure \ref{fig2}, the combination of $\lambda$ and $\xi$ that led to the most robust results (i.e., the smallest variability) was chosen as the default (i.e., $\lambda=2.0$, $\xi=12h^{-1}$Mpc).

We find that the mean number of voids for each pair of parameters ($\lambda$, $\xi$) varies from between 10 to 28 voids. This is a significant difference and is one of the weaknesses of our automatic void finding algorithm. For the optimum choice of parameters, the mean number of voids found is 12. Although we could have chosen parameters which would give us a number of voids more nearly equal to the mean number (all combinations being averaged), we believe that it is more important to balance this with minimal variability. It is worthwhile noting that the absolute number of voids in any sample is relatively uninteresting because it cannot be rigorously defined. However, what is important is that an algorithm be robust (e.g., self-consistent, resistant to variability) so that the \emph{relative differences} in void properties from sample to sample can be reliably determined.

\section{STATISTICAL SIGNIFICANCE OF VOIDS}

The statistical significance of void detection was tested using resampling techniques similar to those used by Kauffmann \& Fairall (1991). This method consists in first finding the number of voids in the original data, and then redistributing the data randomly and finding how often a void is detected as a result of statistical fluctuations. This randomization experiment is repeated many times. This test for the statistical significance of the voids was done for the UZC using the parameters from H\&V's analysis. It was found that 98\% of the time no voids were identified in the randomized data. For our own choice of default parameters, we found that more than 99.5\% of the time no voids were identified. Analogous to the methods used in parametric statistics, we can conclude that the probability of committing a pseudo-Type I error (i.e., claiming to find cosmological voids when in fact the void identification is due to a random fluctuation or a deficiency in the algorithm) is 0.5\%. The equivalent {}``Type II error'' (i.e., finding no voids in a given volume when in fact voids exist) is very difficult, if not impossible, to quantify.

Based on our randomization experiments, we conclude that our void finding algorithm is conservative and thus voids identified by the algorithm are likely to be significant. This is a very important and stringent test since meaningful cosmological inferences can only be made if the void identification is truly reliable.

\section{THE SLOAN DIGITAL SKY SURVEY (SDSS)}

The Sloan Digital Sky Survey is the largest astronomical survey to date. Upon completion, it will have systematically surveyed over one quarter of the sky. The photometric and spectroscopic data is being released periodically. Data Release Five (DR5; Adelman-McCarthy et al. 2007) has been available for download since June 2006. It contains over one million spectra, which are critical to the study of large-scale structures.

The SDSS database offers several possibilities as to the type of magnitude system available and we investigated most of them. The best choice was the Petrosian magnitude system and we adopted it. Petrosian magnitudes are measured using an adjustable aperture which corrects for the fact that more distant galaxies appear smaller in size. For this reason, Petrosian magnitudes are a better measure of the true apparent brightness of a galaxy (Petrosian 1976). In other words, the Petrosian magnitude of a galaxy will be approximately the same regardless of its distance from us.

The SDSS data is measured through five filters (i.e., \emph{u}, \emph{g}, \emph{r}, \emph{i}, \emph{z}). We used absolute magnitudes in the \emph{r}-band because we found that the measured \emph{K}-correction showed less spread than in the other filters (see Figure 15 of Blanton \& Roweis 2007).

We have included reddening and an effective \emph{K}-corrections using the following formula:

\begin{equation}
M_{r}=m_{r}+5-5\log(D_{c}(1+z))-A_{r}-\tilde{K}(z)
\end{equation}

\noindent where $m_{r}$ is the apparent Petrosian magnitude in the $r$-band, $z$ is the redshift, $D_{c}$ is the comoving distance given by Equation (\ref{eq:comovingdist}) and thus $D_{c}(1+z)$ is the luminosity distance, and $A_{r}$ is the Galactic extinction in magnitude units. Precise \emph{K}-corrections are complicated to compute, thus we have used an effective \emph{K}-correction of $\tilde{K}(z)\approx1.05z$ after inspection of figure 15 of Blanton \& Roweis (2007). There is some scatter around this line because \emph{K}-corrections depend on the ensemble averages of the spectral energy distribution (SED) of each individual galaxy. Nonetheless, we believe that this is still a reasonably good approximation for the redshift range for which we are applying the correction. Indeed, the size of the scatter is comparable to the uncertainty in the magnitude $m{}_{r}$ for low redshifts. Moreover, using the words of Blanton et al. (2003), {}``\emph{K}-corrections are inherently uncertain'' even when {}``properly'' computed. In fact, we have redone the statistical analyses (described in the next section) with no \textit{K}-corrections and found that the conclusions were unaffected.

We selected an absolute magnitude and volume limited sample. We used the data located in the volume $140^{\circ}\leq\alpha\leq230^{\circ}$ and $30^{\circ}\leq\delta\leq65^{\circ}$ (2286 square degrees) up to a redshift of 0.16 with absolute magnitude in the $r$-band in the interval $-22.3\leq M_{r}\leq-20.8$. There are a few small unsampled regions or holes in the data and they are treated in the same way as the outer boundaries.

We found a total of 232 voids in the sampled volume. A void catalog describing the properties of these voids can be found at the URL \url{http://physics.ubishops.ca/sdssvoids}. The catalog is a compendium of information containing an up to date list of the properties of each void such as their position, the size of the axes of the best-fit ellipsoid and the equivalent spherical radii. It should be noted that we decided to discuss the properties of only those voids whose center is located at a redshift smaller than 0.16 because of the rapid decrease in the number density of galaxies (especially beyond $z=0.2$).

\begin{figure}
\plotone{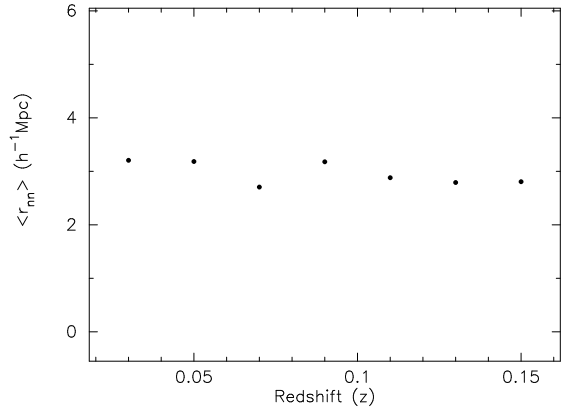} \caption{Distribution of the mean nearest-neighbor distance as a function of redshift for the selected sample. The distribution is overall uniform with a mean nearest-neighbor distance of $\approx3.0$ $h^{-1}$ Mpc. \label{fig3}}
\end{figure}

\section{VOID PROPERTIES}

In order to interpret the properties of the cosmological voids in our sample, the results must be contrasted with theoretical predictions. Large N-body simulations are typically generated in order to understand the formation and evolution of structure in the universe under a given paradigm. According to the $\Lambda$CDM model, voids in \emph{dark matter halo} distributions grow outward from negative density perturbations and tend toward higher and higher sphericity until their boundary reaches that of a neighboring growing underdensity; this moment is known as shell-crossing. If a small underdensity is embedded inside a dense region, it collapses and disappears. Afterward, voids either expand with the Hubble flow or merge to form even larger voids (for more details on the theoretical evolution of voids, see Bond \& Jaffe 1999; Colberg et al. 2005; Dubinski et al. 1993; Sheth \& van de Weygaert 2004). Thus, assuming that galaxies are reasonable tracers of the underlying dark matter distribution, voids in comoving coordinates should either grow or remain approximately constant in size with time (i.e., decreasing redshift) on average as they expand and merge. Given the tremendous wealth of observational data, it is worthwhile to investigate whether or not the theoretical predictions for the distribution and evolution of voids in dark matter distributions can be confirmed by the currently available observational data of galaxy distributions.

\subsection{Void Sizes}

We have attempted to study the variation in void sizes as a function of redshift in order to probe void evolution. Our range of redshift (i.e. $z\le0.16$) corresponds to a look-back time of approximately $t_L=1.36h^{-1}$ Gyrs, which is a relatively small fraction of the history of the universe. Thus we do not expect  to see significant evolution over this redshift range. Nevertheless, this is an interesting pilot study which can easily be followed up as the next generation of large redshift surveys, probing larger volumes and higher redshifts, become available. The main confounding factor in studying the change in sizes of voids with redshift is due to variations in the mean nearest-neighbor distance ($\left\langle r_{nn}\right\rangle$)\footnote{We define $\left\langle r_{nn}\right\rangle=\sum_{i=1}^{n}\left(r_{i}\right)/\sum_{i}^{n}\left(1\right)$ where $n$ is the number of galaxies in a given bin and $r_{i}$ is the distance to the nearest-neighbor for the $i^{th}$ galaxy. Keeping $\left\langle r_{nn}\right\rangle$ constant ensures a homogeneous sampling even if structures such as large voids change the overall number density in a bin.} as a function of redshift (we are assuming a sufficiently high-density, volume-limited sample). We have examined this issue by arbitrarily removing galaxies from samples and we conclude that the sizes of voids increase linearly with increasing $\left\langle r_{nn}\right\rangle$. Indeed, past analyses have shown this strong dependency of the sizes of voids on the density of galaxies (Muller et al. 2000; Muller \& Maulbetsch 2004).

With respect to our sample (see Figure \ref{fig3}), $\left\langle r_{nn}\right\rangle$ is reasonably constant. For this reason, we are confident that any observed change in the average sizes of cosmological voids over the entire range of redshifts is independent of $\left\langle r_{nn}\right\rangle$. We find a roughly constant void volume filling fraction of $81$\% for $0.04 \le z \le 0.16$.

\begin{figure}
\plotone{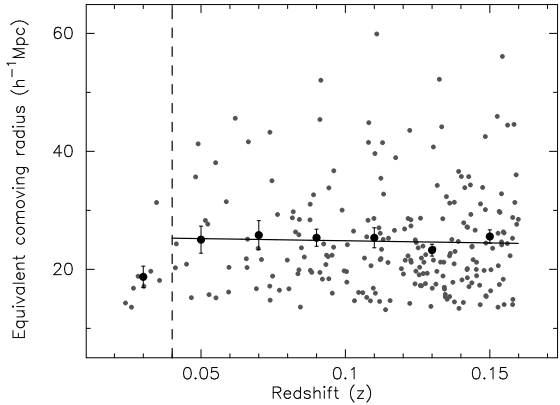} \caption{The equivalent (grey points) and average equivalent comoving radius of voids (black points) as a function of redshift in the SDSS. The error bars show the standard error on the average. The slope of the best fit line (solid line) to the average equivalent radii for $z\le0.04$ is consistent with zero. The data to the left of the vertical dashed line have been removed from this analysis in order to reduce edge effects due to the size and shape of the volume and errors due to cosmic variance. \label{fig4}}
\end{figure}

We group voids in redshift bins of $\bigtriangleup z=0.02$ and obtain the average size of voids for each bin. We decide to exclude the data below $z=0.04$ from our analysis to minimize possible edge effects due to the shape and size of the sampled volume and the effects of cosmic variance. We use a weighted least squares analysis to apply a weight to each point according to its standard error and find that the rate of increase in the average size of voids as a function of redshift is $m=-7\pm20h^{-1}$Mpc per unit redshift (see Figure \ref{fig4}). Thus, the slope is consistent with being null or negative as predicted by theory.

Therefore, while there is a hint that voids may be increasing in size with time (i.e. decreasing redshift) over the limited redshift range probed, the data is statistically consistent with no size evolution. Our sample encompasses 40\% of DR5 or 5\% of the whole sky, therefore it might suffer from cosmic variance especially at low redshit. For this reason, this study could benefit from a larger sample, either covering a more expansive proportion of the sky or various directions.

\subsection{Shape and Orientation}

\begin{figure}
\plotone{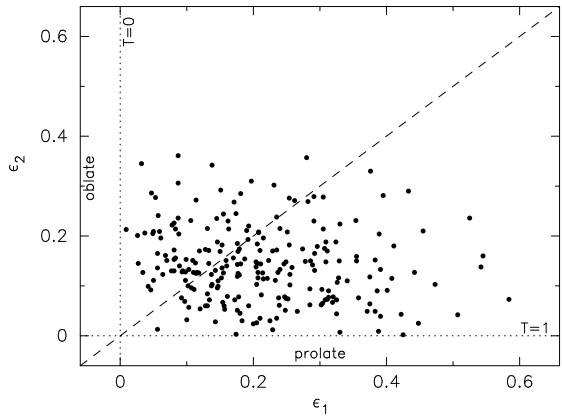} \caption{The distribution of the shapes of the voids in our sample. A slight majority of voids are prolate in shape. \label{fig5}}
\end{figure}

We have also analyzed the distribution of the orientation and shape of our identified cosmological voids. Perhaps not surprisingly, we find a slight majority of prolate voids ($T_{average}=0.67\pm0.09$) in agreement with the theoretical models of Platen et al. (2007). Figure \ref{fig5} shows the bivariate distribution of $\epsilon_{1}$ versus $\epsilon_{2}$.

\begin{figure}
\plotone{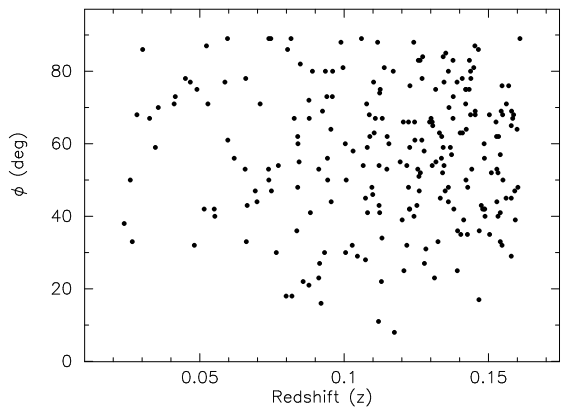} \caption{Orientation of cosmological voids as a function of redshift. There is no significant trend with redshift. The average orientation is $\phi=57.9^{\circ}$. \label{fig6}}
\end{figure}

Following the suggestion of Ryden \& Mellot (1996, see Section \ref{sec:INTRODUCTION}), we quantify the orientation of cosmological voids with respect to the line-of-sight as a measure of the effects of redshift-space distortions on our identified cosmological voids. We find an average orientation with respect to the line-of-sight of $\left\langle\phi\right\rangle= 57.9^{\circ}$. In order to determine whether or not the distribution of our measured values of $\phi$ is consistent with being drawn from a random distribution, we used a bootstrap resampling test. We took 100,000 samples of 222 voids from our sample for $0.04 \le z \le 0.16$ by sampling with replacement. The 99\% confidence interval on the average orientation was: $54.6^{\circ} < \left\langle\phi\right\rangle < 61.2^{\circ}$. A similar analysis was performed for the median ($\tilde\phi$) of the population. The 99\% confidence interval on the median was: $53.6^{\circ}<\tilde\phi<65.5^{\circ}$. This is consistent with an isotropic distribution of orientations for which we would expect the $\tilde\phi$ to be 60$^{\circ}$ and $\left\langle\phi\right\rangle$ to be 57.3$^{\circ}$. We also looked at the orientation of voids in redshift slices of thickness $\Delta z=0.01$. We conclude that the scatter in each slice is consistent with random statistical fluctuations in the orientations within each slice. However, there is the suggestion that redshift-space distortions may be causing the voids at redshifts of approximately $0.10$ (on average lower values of $\phi$) to have increasingly higher values of $\phi$ as $z\to0.16$. Indeed, it can be seen in Figure \ref{fig6} that there are relatively fewer voids at high redshifts with low values of $\phi$ compared to $z=0.1$. This result is sufficiently interesting that larger portions of the sky need to be analyzed before a definitive statement can be made.

There are several caveats that should be noted with respect to the analysis of the orientations of voids and their connection to redshift-space distortions. First, as with the above radius analysis, this result may be affected by cosmic variance, especially at low redshift. Second, using N-body simulations, Platen et al. (2008) have shown that the $\Lambda$CDM model predicts that the orientation of voids in the distribution of dark matter haloes can be correlated for scales $<30h^{-1}Mpc$. On the other hand, they also find that this predicted void alignment becomes less significant for larger scales. They demonstrate that the alignment is likely related to the cosmic tidal field. While our sample reaches well beyond this scale, we cannot rule out the effects of tidal alignment on our results.

\subsection{Comparison to $\Lambda$CDM}

In order to test whether or not our results are consistent with the predictions of the $\Lambda$CDM model, we perform an identical analysis on a mock SDSS catalog lightcone of Croton et al. (2006). The mock catalog is obtained from the Millennium Simulation via the use of the semi-analytic model of Croton et al. (2006) with the dust prescription of Kitzbichler \& White (2007). We use the ``observed'' redshift (i.e., including peculiar velocities) in order to directly compare with our observational analysis of SDSS. We find that the average distance to the nearest neighbor (i.e. the equivalent of Figure \ref{fig3}) is also constant, albeit at roughly $2.5 h^{-1}$ Mpc, for $z\lesssim0.14$. Moreover, the magnitude limited luminosity function is in good agreement with the SDSS dataset for redshifts of $0.06\le z \le0.12$. This ensures that the results obtained from running our void finding algorithm on the mock catalog are comparable to those of the SDSS catalog.

\begin{figure}
\plotone{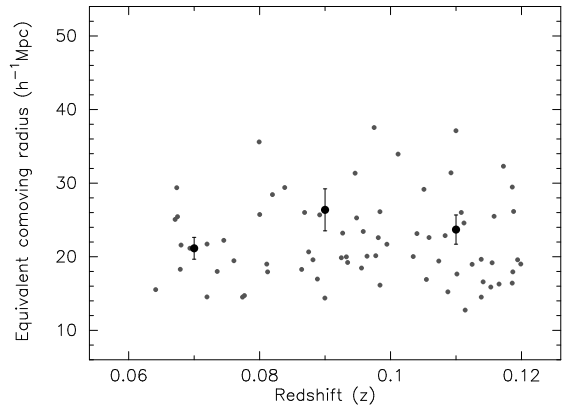} \caption{The equivalent (grey points) and average equivalent comoving radius of voids (black points) as a function of redshift in the \emph{mock} SDSS catalog. The error bars show the standard error on the average. \label{fig7}}
\end{figure}

While the redshift range is slightly smaller than for the SDSS survey, we mostly find reasonable agreement between the theoretical and observational analyses. Indeed, the equivalent comoving radii of cosmological voids agree well (see Figure \ref{fig7}). 

\begin{figure}
\plotone{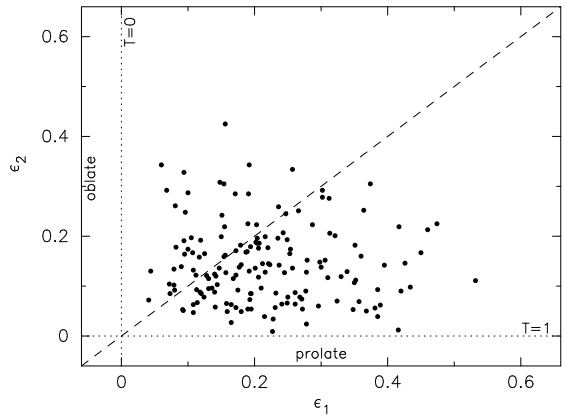} \caption{The distribution of the shapes of the voids in our mock galaxy sample. As with the SDSS catalog, a slight majority of voids are prolate in shape. \label{fig8}}
\end{figure}

According to our definition of voids, the void volume filling fraction for both the mock catalog and the SDSS sample for the same redshift range is approximately 80\%. Specifically, we find a void volume filling fraction to be $77$\% for the mock catalog and $83$\% for the SDSS sample ($0.06\le z\le0.12$). We also find a slight majority of prolate voids with an average triaxiality parameter of $T_{average}=0.65\pm0.09$ in good agreement with the SDSS results. This can be seen in Figure \ref{fig8} where we show the bivariate distribution of $\epsilon_{2}$ versus $\epsilon_{1}$.

In contrast to the SDSS void sample, the 99\% confidence interval for the average orientation of cosmological voids in the mock catalog is $61.3^{\circ} \le \left\langle\phi\right\rangle \le 69.3^{\circ}$. The results strongly indicate (with 99.99\% confidence) that the voids in the mock catalog are on average compressed along the line-of-sight (i.e., $\phi > 60^{\circ}$). A similar conclusion that $\tilde\phi$ is significantly greater than the value for random orientations (i.e., $60^{\circ}$) is also evident.

In order to quantify whether the void orientations from the mock catalog are different from those of our SDSS sample over the same redshift range, we use a bootstrapping method to evaluate the probability that the two samples are derived from the same population. We find that the probability that the two samples have the same values of $\left\langle\phi\right\rangle$ is less than 0.2\% and that they have the same values of $\tilde\phi$ is less than 0.1\%. Provided that the two samples are indeed comparable, this discrepancy between our analysis of the mock galaxy catalog and that of SDSS could be reconciled through fine-tuning of the initial conditions of the assumed cosmology or of the input semi-analytic prescriptions. However, the details of the required fine-tuning are non-trivial and beyond the scope of this paper.

\section{CONCLUSIONS}
Cosmological voids have only been rigorously studied for the past 20 years. Much effort has been put into trying to find out if cosmological parameters can be extracted from the study of large-scale structures. Our primary focus is on the quantification of the properties of cosmological voids. In order to accomplish this, we look at several void finding algorithms and choose to implement an algorithm similar to that of H\&V. Our algorithm is designed to objectively identify voids in a galaxy distribution in a manner that would best reproduce the results from a visual inspection.

The void finding algorithm is thoroughly tested using the UZC and the PSCz survey, and our results are compared with those of H\&V. In general, good agreement is found between the two methods. Our algorithm proves to be very conservative which guarantees with a high probability that the identified cosmological voids are not spurious.

Using the results from DR5 of the SDSS, we carefully identify regions where the survey data is nearly spatially contiguous and use that data to identify voids. Our galaxy sample is selected in the region corresponding to $140^{\circ}\leq\alpha\leq230^{\circ}$ and $30^{\circ}\leq\delta\leq65^{\circ}$ up to a redshift of $z=0.16$. We then fix the absolute magnitude limits to obtain a homogeneously sampled dataset of galaxies.

We repeat our analysis on a mock SDSS catalog based on the semi-analytic models of Croton et al. (2006). The magnitude limited luminosity function and the average distance to the nearest neighbor in the galaxy sample derived from the mock catalog show good agreement with our SDSS sample for redshifts $0.06 \le z \le 0.12$. This guarantees that the results from both analyses are comparable.

The observational (SDSS) and theoretical (mock catalog) analyses yield similar void sizes and void volume filling fractions. We find no statistically significant trend between the average size of cosmological voids and redshift for $0.04\le z\le0.16$ in the SDSS sample.  We obtain a slight majority of prolate voids in both the observational and theoretical dataset.

Following the suggestion of Ryden \& Mellot (1996), we study the distribution of the orientation of cosmological voids with respect to the line-of-sight in redshift space as a probe of redshift-space distortions. We find that the orientations of cosmological voids in the SDSS sample with respect to the line-of-sight are consistent with being randomly distributed, indicating that redshift-space distortions are either not present or too weak to be detected with the present analysis. Indeed, both the median and average orientation of cosmological voids in the SDSS are consistent with being drawn from an isotropic distribution.

In contrast, the orientation of voids in the mock catalog tend to be compressed along the line-of-sight. Using bootstrapping methods, we find a 0.1-0.2\% probability that the distribution of the orientation of cosmological voids over a similar redshift range in both samples are consistent with having been drawn from the same parent population. 
Assuming that this difference in the alignment of voids with respect to the line-of-sight between the mock and the SDSS data is real and due to redshift-space distortions, which are influenced by the initial conditions of the universe in a non-trivial manner, studies such as ours could in principle help better constrain the cosmological parameters and semi-analytic models. While beyond the scope of the present project, the same analysis could be performed on mock catalogs from cosmological simulations of varying initial conditions and input semi-analytic model prescriptions.

Finally, given the existence of other void finding algorithms that use a wide variety of techniques to identify voids (see Colberg et al. 2008), this analysis could be repeated using a totally different set of criteria for void identification and thus the validity of the measured properties can be independently verified. This would help constrain the possibility of systematics introduced by our void finding method. 

\acknowledgments

C. Foster would like to thank NSERC (Canada) for support in the form of a graduate scholarship. L. Nelson would like to thank the Canada Research Chairs Program and NSERC for financial support. L. Nelson would also like to acknowledge the support of the Kavli Institute for Theoretical Physics where part of this work was completed. We also thank M. Vogeley, D. Bond, C. Blake, R. Crain, R. van Hulst and M. Zemp for helpful discussions. We acknowledge the help of D. Croton in providing both the mock catalog data and related support. Finally, we are especially grateful to the anonymous referee for his/her careful reading of this paper and for all the helpful comments that helped to improve it.

\end{document}